\documentclass[conference]{IEEEtran}

\usepackage{dsfont}      
\usepackage{amssymb}      

\usepackage{amsmath}
\usepackage{graphicx}
\usepackage{subfigure}
\ifCLASSINFOpdf
\else
\fi
\begin{document}
%
\title{Does Your DNS Recursion Really Time Out as Intended? A Timeout Vulnerability of DNS Recursive Servers}



\author{\IEEEauthorblockN{Zheng Wang}
\IEEEauthorblockA{Information Technology Laboratary\\
National Institute of Standards and Technology\\
Gaithersburg, MD 20899, USA\\
Email: zhengwang98@gmail.com}
}


%


\maketitle

\begin{abstract}
Parallelization is featured by DNS recursive servers to do time-consuming recursions on behalf on clients.
As common DNS configurations, recursive servers should allow a reasonable timeout for each recursion which may take as long as several seconds.
However, it is proposed in this paper that recursion parallelization may be exploited by attackers to compromise the recursion timeout mechanism for the purpose of DoS or DDoS attacks.
Attackers can have recursive servers drop early existing recursions in service by saturating recursion parallelization.
The key of the proposed attack model is to reliably prolong service times for any attacking queries. As means of prolong service times, serval techniques are proposed to effectively
avoiding cache hit and prolonging overall latency of external DNS lookups respectively. The impacts of saturated recursion parallelization on timeout are analytically provided.
The testing on BIND servers demonstrates that with carefully crafted queries, an attacker can use a low or moderate level of query load to
successfully overwhelm a target recursive server from serving the legitimate clients.
\end{abstract}


%
\IEEEpeerreviewmaketitle

\section{Introduction}
The Domain Name System (DNS) plays a critical role in today's Internet. It translate human friendly domain names into machine readable IP addresses, which is indispensable for Internet addressing and routing. 
The design of DNS basically follows the client-server model. The server side consists of recursive server and authoritative server. Recursive server performs iterative name resolution on behalf of client. 
And authoritative server provides authoritative response to queriers.

As an DNS agent between DNS user and DNS database, recursive server processes user requests and traverses the DNS tree to reach final answer. 
Given its heavy tasks and complicated interoperations, the design space of recursive server is very complex, with many optimization proposals emerging in recent years [1]-[8].
Its performance and security are of great importance to the stability, resilience, and security of the DNS as a whole [9]-[22]. There have been extensive studies on the securities of recursive server [28]-[41] in the past decade. However, little attention was given to the parallelization characterized by recursive server. 
This paper is the first to explore the security implications of parallelization in recursive server. A new vulnerability is discovered to potentially cause Denial of Service on recursive server.   

\section{Parallelization Featured by Recursive Servers}

Without reliance on external DNS lookups, an authoritative server generates responses based on the local DNS information,
namely the authoritative zone stored locally. So the query processing delay of an authoritative server is only
dependent of its processing capability (whose bottleneck is affected by CPU, memory, hard drive, etc.).
It usually takes no more than the order of microsecond to process a query for an authoritative server.

With the responsibility of finding the final answer to a recursion-desired query, a recursive server may look up external
authoritative servers following the referrals obtained by far. The recursive process may cost several RTTs when multiple zone
cuts take place from the nearest domain in cache to the target domain. So the query processing delay of a recursive server is
mostly determined by the external lookups involved in the recursive resolution \cite{Impact}. For example, if two external lookups are
needed and each costs 200 ms,  the total query processing delay is about 400ms. Unlike authoritative servers, the internal
processing delay is negligible for recursive servers.

Given the big difference in query processing delay between authoritative servers and recursive servers,
the parallelization is totally different between them. Authoritative servers do not, in essence, need high parallelization
in handling the incoming queries. They can process queries in a FCFS queue. By contrary, recursive servers are often dominated
by parallelization in processing queries. A number of concurrent queries should be processed in parallel by recursive servers.

\section{The Configuration of Recursion Parallelization and Timeout in BIND and Other DNS Servers}

Each recursion, of course, consumes an mount of memory. To limit the memory usage of parallel recursions,
BIND allows the configuration of maximum recursions. The default is 1000.

Some recursions may take unreasonably long times. To cut the time waiting for the excessively delayed responses,
BIND allows to configure the amount of time the recursive resolver will spend attempting to resolve a recursive
query before failing. The default and minimum is 10 and the maximum is 30.

Some other DNS servers also provide their users the opportunity of configuring recursion parallelization and timeout to optimize
performance and ensure reliability.

\section{Timeout Impacted by the Saturated Recursion Parallelization}
In BIND, recursive-clients defines a "hard quota" limit for pending recursive clients: when more clients
than this are pending, new incoming requests will not be accepted, and for each incoming request a previous
pending request will also be dropped. A "soft quota" is also set. When this lower quota is exceeded, incoming
requests are accepted, but for each one, a pending request will be dropped. If recursive-clients is greater than
1000, the soft quota is set to recursive-clients minus 100; otherwise it is set to 90\% of recursive-clients.
A recursion may be dropped before it really times out when the lower quota is exceeded.

If the attacker can manage to accumulate the parallel recursions to the lower quota, the recursive service will
be subject to a Denial-of-Service (DoS) or Distributed Denial-of-Service (DDoS) attack. Unlike the type of DoS or DDoS attacks on recursive servers
which leverage cache poisoning to inject bogus data into responses \cite{Revisit, Poster}, the proposed new attack model can never be
alleviated by DNSSEC.

\section{Quantifying the Impacts}

Let the limit of recursive clients (recursive-clients) be $L$, the maximum inbound query rate be $Q$,
and the minimum real query timeout be $T$.

The impacts of saturated recursion parallelization on timeout can be expressed as
\begin{equation}
L= Q*T
\end{equation}

If each recursive client consumes $U$ unit of memory, we have the minimum memory requirement for the recursive server as
\begin{equation}
M=L*U
\end{equation}

\textbf{Example 1.}
Let the recursive-clients be the default (1000), the query rate issued by the attacker be 1000 qps. Then the real query timeout will be 1 s, which is much shorter than the default (10 s). It is not uncommon for a DNS lookup to cost more than 500 ms. Then if a recursion takes more than two DNS lookups, it may be dropped early before it times out as intended.

\textbf{Example 2.}
What if the query rate is 10,000 qps? That query rate is usually not seen as a DDoS attack. The real query timeout will be 100 ms. Many recursions will be lost due to the early discarding.

\textbf{Example 3.}
Let the effective query timeout and recursive-clients be both the defaults, namely 10s and 1000 respectively, the maximum attacking query rate will be as small as 100 qps. So the expected query timeout is readily to fail.

\textbf{Example 4.}
BIND is reported to need about 20KB memory usage per recursive client. Let the effective query timeout be the default, namely 10s, and the maximum inbound query rate allowed (or not limited) be 100k qps. Then the minimum memory required for the recursive service will be 10*100k*20k=20GB.


\section{Service Time Matters for the Attacks}

However, the attacking queries do not necessarily take effect if they cannot be served for a sufficient time. They should at least result in a service time of real query timeout.
We can calculate the minimum service times for the four examples discussed above.

\textbf{Example 1.}
The minimum service time is 1s.

\textbf{Example 2.}
The minimum service time is 100 ms.

\textbf{Example 3.}
The minimum service time is 10 s.

\textbf{Example 4.}
The minimum service time is 10 s.

\section{How to Reliably Prolong the Service Time?}

Most recursions will either hit the cache or end with a limited number of  prompt external DNS lookups.
However, there are still several techniques to achieve a reliably long service time.

\subsection{Avoiding Cache Hit}

Cache hit is, of course, unfavorable for a long service time, since the response can be generated immediately from the cache.
So the attacker has to use some techniques to get around the caching.

\subsubsection{Randomizing query names} DNS caching mechanism searches for the item in cache that exactly matches the query name in question.
So if query names may be sufficiently randomized, there is little chance of repeated query names within one TTL and thus the cache hit rate will be close to zero \cite{Flooding} (illustrated in the question in blue in Fig. 1).

\subsubsection{Requesting for authoritative records with a zero TTL} DNS caching mechanism keeps the data in cache for a duration that specified in the TTL field of the relevant record.
So when the TTL is set zero, the associated record is just used for responding once and then dropped immediately (illustrated in the question in red in Fig. 1). The attack can use any existent authoritative records on the DNS name space with a zero TTL.
To ensure the constant availability of and the reliable access to such records, the attack can also create some authoritative records with a zero TTL, or even set up a dedicated DNS zone with records of a zero TTL.

\begin{figure}[!t]
\centering
\includegraphics[width=3.2in]{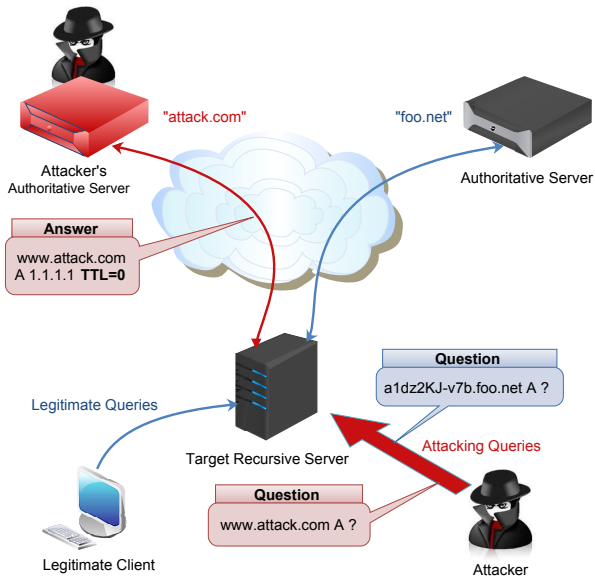}
\caption{Techniques of avoiding cache hit.}
\label{T00}
\end{figure}

\subsection{Increasing Overall Latency of External DNS Lookups}

In most cases, avoiding cache hit alone cannot guarantee a long service time,
because most DNS authoritative servers provide response delay as low as no more than 50 ms in a bid to optimize user perceived performance.
Here we propose two means of prolonging overall latency of external DNS lookups.

\subsubsection{Causing many external DNS lookups} If one external DNS lookup is not enough, many a lookup makes a long service time.
Using a long CNAME/DNAME chain or using query names with deep labels helps to produce many external DNS lookups.

\subsubsection{Requesting the slowly responding domains} A slowly responding domain is surely able to add to the service time. There are two ways of making slowly responding domain available.

\begin{itemize}
  \item Finding slowly responding domains on the Internet.
  The method is to test some candidate domains in operation, and filter out those with relatively quick responses.
  In order to measure the real response time between the target recursive server and the authoritative server,
  the test procedure can be detailed as follows: \\
 \emph{ 1) Send a request for a testing domain to the target recursive server and get the response (Step 1 and 2 in Fig. 2).\\2)
  Check the TTL in the response. And within that TTL, send the same request again to the target recursive server (Step 3 in Fig. 2).
  Measure the RTT of that response. And that RTT is the estimation of the RTT between the client (attacker) and the target recursive server,
  which is denoted by $R_c$. The first request intends to make the response cached by the recursive server and the second request
  intends to get the response immediately from the cache.\\
  3) Send a request for a candidate domain to the target recursive server and get the response.\\
  4) Check the TTL of the authoritative zone's NS RRset in the response. And within that TTL, send another request for a randomized domain within
  that authoritative zone to the target recursive server (Step 4 in Fig. 2). The second request is expected to miss the cache and trigger an
  outstanding query for the authoritative zone. Measure the RTT of the second response. And that RTT is the RTT between the client
  (attacker) and the target recursive server plus the RTT between the target recursive server and the authoritative servers, which is denoted by $R_t$.
  \\5) Finally, the RTT between the target recursive server and the authoritative servers can be estimated by $R_a=R_t-R_c$.\\
  Since one authoritative zone may have multiple authoritative servers and the target
  recursive server may select one server from them to query each time. Most server selection
  algorithms make every server have a chance to be selected \cite{ServerSelection}.\\
  To ensure all authoritative servers for a candidate domain are located far enough from the target recursive server,
  that candidate domain should be measured enough times, but, of course, using different randomized domains.
  So each authoritative server should have an opportunity to be measured. And By checking if all measurements for a domain exceed the minimum delay, we can ensure the slow responses from that domain.\\}
  \item Setting up malicious slowly responding domains. To facilitate the attack, the attacker can host malicious slowly responding domains
  which are targets of the attacking queries.
\end{itemize}

\begin{figure}[!t]
\centering
\includegraphics[width=3.2in]{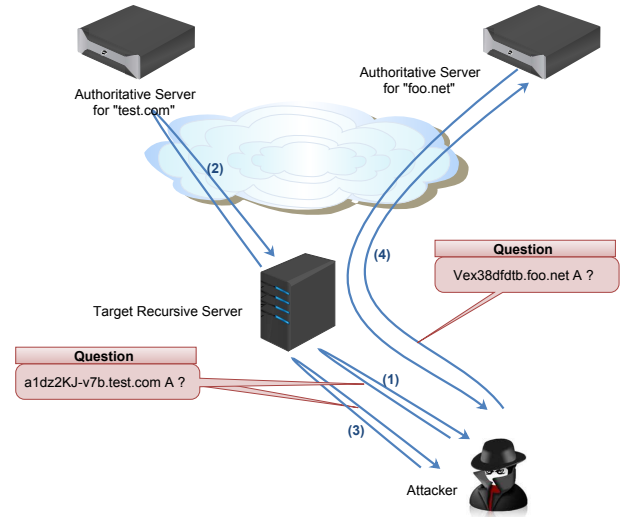}
\caption{Procedure of measuring response delay.}
\label{T00}
\end{figure}

\section{Testing}

We configured a BND recursive server (BIND 9.10.4-P1, which is the most recent stable version of BIND) with the default recursive-clients (1000) and timeout (10s).

"auditionsea.com" was selected as the requested domain of attacking queries. It was tested to have response delay of more than 500 ms using the measurement technique detailed above.
We used resperf to generate a query rate of 5000 qps for randomized query names towards the recursive server.

At the same time, we used a client as a legitimate client to generate a query rate of 1 qps towards the recursive server. We select "whois.net" as the requested domain of that legitimate client. It was tested to have response delay of more than 200 ms using the measurement technique detailed above.
Since both "auditionsea.com" and "whois.net" satisfy the minimum service time required per Eq.(1), they are expected to be dropped early by the recursive server under the attack as we discussed above.

After the initial bootstrap, all queries, for both "auditionsea.com" and "whois.net", were observed to receive the SERVFAIL responses from the recursive server.
Those results meet our expectations. And it demonstrates that the attacker can use a low or moderate level of query load to
successfully overwhelm the target recursive server from serving the legitimate clients.

To better understand the testing results, we captured the query and response trace for further analysis. We find that all queries indeed got around the cache and triggered external queries as we expected, since the outgoing queries matching the incoming queries were observed from the recursive server.
We also find that those outbound queries were successfully responded with NXDOMAIN by the authoritative servers because of their randomized query names. However, those responses were observed to arrive at the recursive server only to be discarded, because
because there were no longer outstanding processes pending for them.

\section{Conclusions}
Parallelization is indispensable for DNS recursive servers to maintain good throughput. However, this paper proposed that parallelization also imposes DoS or DDoS risks for recursive servers.
In the proposed attack model, carefully crafted DNS queries are employed to saturate parallelization of recursive servers and thereby persistently overwhelm the recursive services.
The testing on BIND validated the proposed attack model.
%
%



%

\end{document}